# Ferroelastic switching with van der Waals direction transformation in layered PdSe$_2$ driven by uniaxial and shear strain

Peng Lv[1], Gang Tang[1], Yanyu Liu[1], Yingzhuo Lun[1], Xueyun Wang[1] and Jiawang Hong[1*]

[1]*School of Aerospace Engineering, Beijing Institute of Technology, Beijing 100081, China*

## Abstract

Uniaxial and biaxial strain approaches are usually implemented to switch the ferroelastic states, which play a key role in the application of the ferroics and shape memory materials. In this work, by using the first-principles calculations, we found not only uniaxial strain, but also shear strain can induce a novel ferroelastic switching, in which the van der Waals (vdW) layered direction rotates with the ferroelastic transition in layered bulk PdSe$_2$. The shear strain induces ferroelastic switching with three times amplitude smaller than uniaxial strain. The novel three-states ferroelastic switching in layered PdSe$_2$ also occurs under shear strain. Our result shows that the shear strain could be used as an effective approach for manipulating the functionalities of layered materials in potential device applications.

[*] Corresponding author. E-mail: hongjw@bit.edu.cn (J. Hong);

# 1. Introduction

As one of the mechanical-related coupling order parameters, ferroelasticity plays a key role in the application of the ferroics [1-3] and shape memory materials [4-6]. Nowadays, uniaxial and biaxial strain approaches have been widely employed to achieve the ferroelastic switching [7-9]. Since these homogeneous tensile/compressive strain usually does not break the mirror symmetry of crystal structure for specific structural distortions, thus preventing people from realizing some orientated states [5]. However, the shear strain can break the mirror symmetry [5, 10] and it will be interesting to know if the shear strain can induce the two equivalent variants. Yet the shear strain has been rarely employed to investigate its effect on the van der Waals (vdW) layered ferroelastic materials.

Traditionally, based on the characteristics of the shear strain itself, many researchers employed the shear strain to investigate the failure mechanisms with ideal shear strengths [11-17], and the exfoliation of 2D materials [18-20]. More recently, it has been found that shear strain has novel applications in many aspects of material research [21-23], causing unexpected novel phenomena, such as funnel effect of excitons [24] and dominant channel of spin admixture [25]. Very recently, it found that the shear strain can act as an ultrafast symmetry switch to control the topological property of $WTe_2$ Weyl semimetal by deforming the orthorhombic structure into the monoclinic phase [26]. It was showing that shear strain provides a more efficient means to manipulate the topological properties than uniaxial strain. Under these circumstances, there is a strong impetus to employ shear strain to explore more special and incredible issues involved symmetry for novel materials, including ferroelasticity.

Orthorhombic $PdSe_2$ is a novel layered material with an in-plane pentagonal structure, which is different from the typical transition metal dichalcogenides [27]. The unique atomic structure and strong interlayer coupling offer novel electronic properties and good air stability for nanoelectronics [28], as well as effortless defect hopping in both intralayer and interlayer [29]. $PdSe_2$ has been widely investigated and attracted intensive attentions due to its novel properties which are different from those of commonly known 2D materials [30-32]. Very recently, it was reported that ferroelastic switching occurs in $PdSe_2$ with the uniaxial compressive stress along vdW direction [33]. However, it is not clear if the strain along in-plane direction can still induce ferroelastic switching. Meanwhile, it is also very interesting to explore how the shear strain affect its mechanical and electric behaviors.

In our work, we found that not only compressive strain along $c$ direction (vdW layered direction) but also uniaxial tensile strain both along $a$ or $b$ directions (in-plane) can induce the ferroelastic switching in layered materials $PdSe_2$ from density functional theory (DFT) calculations. Furthermore, we found that the shear strain can also cause the ferroelastic switching in $PdSe_2$ and transform the non-vdW direction to vdW direction, leading to a lattice rotation and semiconductor-metal-semiconductor transition, with three times smaller magnitude than uniaxial strain. More interestingly, the shear strain, rather than uniaxial strain, can induce three different ferroelectric states in the layered materials $PdSe_2$ which can be used as three-states ferroelastic switching[34]. It reveals that the ferroelastic phase transition with vdW layer direction change results from the relatively strong interlayer coupling which is usually very weak in layered materials with vdW interactions. Our work suggests that shear strain is an effective way to tune the vdW layer direction through ferroelastic phase transition and bring novel functionalities and potential applications in 2D materials.

## 2. Methods

The generalized gradient approximation (GGA) method of Perdewe–Burkee–Ernzerh (PBE) [35] were performed based on density functional theory (DFT) implemented in the Vienna ab initio Simulation Package (VASP) [36, 37]. The cutoff energy for the plane wave basis set was tested and taken as 450 eV. Both lattice constants and atomic positions were relaxed until the forces on atoms were less than $10^{-3}$ eV/Å, and the total energy change was less than $10^{-7}$ eV. The Brillouin zone integrations were performed by using a $\Gamma$-centered grid of $5 \times 5 \times 4$. The phonon dispersions and second order force constant were calculated with the finite displacement method by using the PHONOPY code [38]. The crystal orbital Hamilton population (-COHP) analysis was performed using the LOBSTER code [39]. Several vdW methods were considered for layered $PdSe_2$ and we found that optPBE functional [40] yields the most reasonable structural lattice constants. More information can be found in supporting materials. To accurately match the experimental result of band gap (0.06 eV) from dI/dV spectrum[29], the electric structure calculations are used by GGA+U (U=4.3 eV) method.

## 3. Results

3.1 Uniaxial strain induced ferroelastic switching

As we all know, applying electric displacement ($D$)[41] to ferroelectric materials enables one to reach the intermediate states during the ferroelectric switching, while applying electric field ($E$) can only have two positive and negative polarization state, without intermediate states [42-44]. This is because the energy landscape is multivalued and the paraelectric configuration is unstable at small $E$ field in a ferroelectric material, while at constant D field, the energy landscape remains single-valued, thus allowing access to the entire electric equation of state for ferroelectric materials[42]. Analogously, for ferroelastic switching, applying strain field ($\varepsilon$) rather than stress field ($\sigma$) will enable us to obtain any intermediate state during the ferroelastic transition. Therefore, it is expected that more states may be obtained with strain field than with stress field during ferroelastic phase transition[33].

Firstly, we obtained the energy-strain relationship curves with uniaxial strain along *a*, *b* and *c* directions, as shown in **Figures 1a** and **1b**. For all three directions, it can be found that there are five special inflection points I, II, III, IV and I', corresponding to the zero stress states in **Figures 1c** and **d**. Among these five special states in **Figures 1a** and **b**, states I, I' and III are stable states since they have positive second derivative, states IV are meta-stable states with negative second derivative in energy profiles (The lattice constants are listed in **Table S2**). We examined the atomic structures (**Figures S1-S3**) of I and I′ states with two identical energy. For the case of strain along *a* direction, we found that these two stable states have the exactly same atomic structure (**Figure S1**) accompanying with 90° lattice rotation, *i.e.*, vdW layered direction along *c* for strain-free structure (state I) changes to *a* direction of strained structure (state I').

Thus, the two lowest energy states (I and I') indeed correspond to two different ferroelastic orientated states under external strain. The similar ferroelastic switching with phase transition can also be found under uniaxial strain along *b* and *c* directions (**Figures S2 and S3**). Also, the stable state III shows nearly equivalent lattice constants of three directions. From this structure, we relaxed it by constraining its lattice constant to be equal and obtained the cubic phase of PdSe$_2$ ($a=b=c=6.32$ Å in our calculation, space group of *Pa*-3). The appearance of meta-stable phase below Curie temperature indicates a *first-order* for the ferroelastic-paraelastic phase transition of PdSe$_2$ [45].

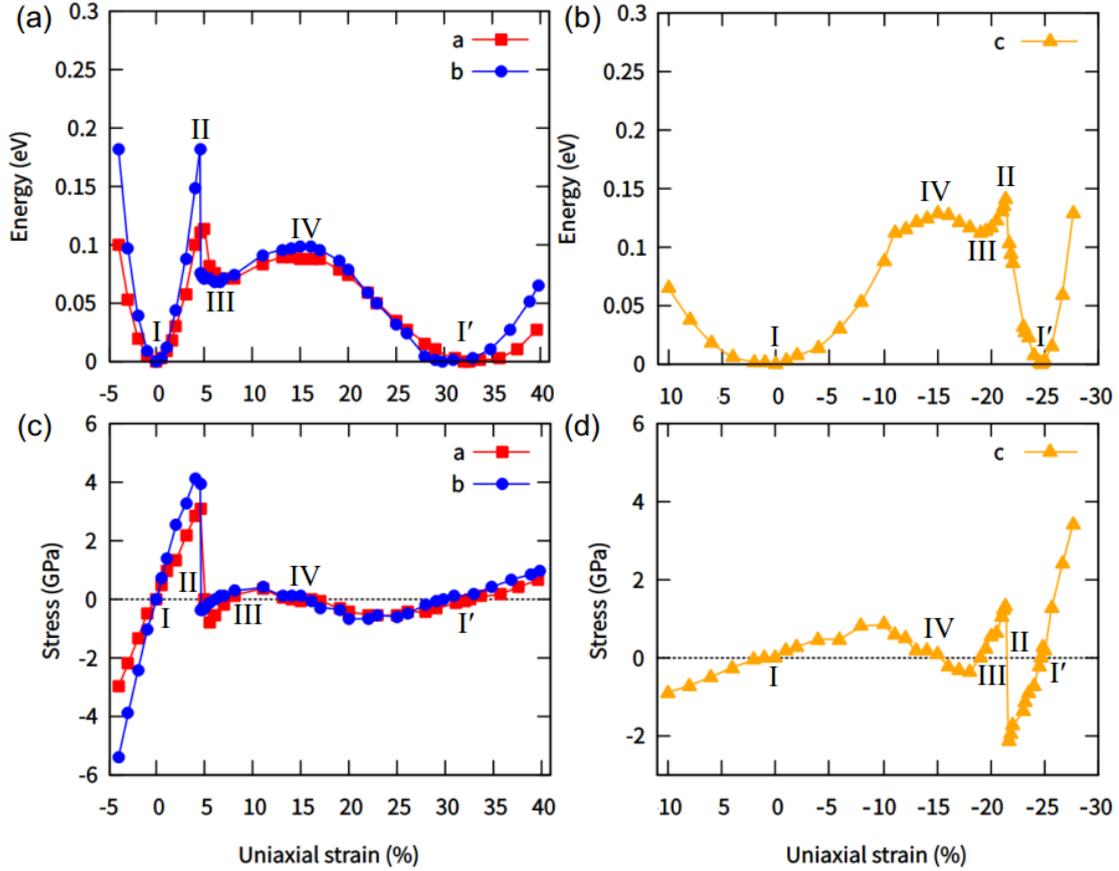

**Figure 1**. The energy-strain relationship under uniaxial strain along (a) *a*, *b* and (b) *c* directions for bulk PdSe$_2$, respectively. The corresponding stress-strain relationship under uniaxial strain are shown in (c) and (d). The positive strain along *a* and *b* directions indicates the tensile deformation and the negative strain indicates the compressive deformation.

We obtained the stress-strain curves (**Figures 1c** and **1d**) and it can be found that the value of stress fluctuates around zero stress which also indicates there are several stable states and phase transitions during the strain loading. From stress-energy curves with strain loading (**Figure S4**), it is clear to show that stress indeed has multiple-valued energy states. Interestingly, comparing the mechanical behaviors between in-plane loading (*i.e.* along *a* or *b* direction) and out-of-plane loading (*i.e.* along *c* direction), it can be seen that both the energy-strain (**Figure 1b**) and stress-strain (**Figure 1d**) curves for out-of-plane loading look "flipped" from those of in-plane loading case (**Figures 1a** and **1c**). This "flipped" mechanical behavior with in-plane and out-of-plane strain loading results from the equivalent ferroelastic transition. For example, with large enough strain along *a* direction (**Figures 1a**), the ferroelastic phase transition occurs in pristine structure (state I) and the new structure has the van der Waals layer oriented along this loading direction (state I'). This state I' in **Figure 1a** is as same as the state I in **Figure 1b** in which the strain loading is also along vdW direction (*c* direction). Therefore, they should show the same behavior for state I' in **Figure 1a** (**Figure 1c**) and state I in **Figure 1b** (**Figure 1d**). Analogously, the state I in **Figure 1a** (**Figure 1c**) also has the same mechanical behavior as state I' in **Figure 1b** (**Figure 1d**). During the ferroelastic switching, it is accompanied by the semiconductor-metal-semiconductor transition (**Figure S5**), not only with mechanical loading along vdW direction, but also along non-vdW direction (*b* and *c* directions). It can be seen that when the strain is closed to the critical strain trigging the ferroelastic state, the band gap can reopen and increase with the strain.

3.2 Shear strain induced ferroelastic switching

Next, we show that shear strain can also induce the ferroelastic switching of bulk $PdSe_2$ with vdW direction transformation. **Figure 2a** shows the shear stress-strain relationship in three shear plane. The directional shear modulus (**Table S3**) can be obtained by fitting the shear stress versus strain with the strain less than 2%. It can be seen that the shear stress monotonically increases until shear strain reaches at 17% (18%) in *ac* (*bc*) slip plane and begins to gradually decreases. The stress trend in *ac* and *bc* slip planes is the typical interlayer slippage, which is similar to that of layered SnSe [46]. However, surprisingly, the *ab* plane shear stress increases much faster than those in *ac* and *bc* planes below 7% and then it suddenly drops at 7%. Very interestingly, the *ab* plane shear stress doesn't decrease to zero, instead, it drops to the stress-strain curve of *bc* plane and then nearly follows its profile, as seen in **Figure 2a**. This abrupt change behavior does not mean mechanical failure, as we checked its structural stability from the phonon dispersion with 7% shear strain, in which no imaginary frequency appears for $PdSe_2$ structure under shear strain of 7% (before sudden change) and 8% (after sudden change), as shown in **Figure S6**. In addition, the stress (**Figure 2a**) and energy (**Figure 2b**) profiles clearly show that there is first-order phase transition occurs near 8% shear straion in *ab* plane.

Due to symmetry breaking of shear strain, the space group of bulk $PdSe_2$ becomes monoclinic $P12_1/c1$ (No. 14) from original orthorhombic *Pbca* (No. 61) with vdW layered direction transformation from *c* direction to *b* direction at 8% *ab* plane shear strain, as shown in **Figure 3a**. After unloading the shear strain, we found the structure recovers to the original phase with space group of *Pbca* (No. 61) but with 90° lattice rotation, *i.e.*, *a*=*b'*, *b*=*c'* and *c*=*a'*. This indicates the *ab* plane shear strain can induce ferroelastic phase transition in layered $PdSe_2$, similar to the uniaxial strain case as we discussed in previous section. However, only *ab* plane shear strain can induce this ferroelastic phase transition, because other two shear strains in *bc* and *ac* planes only induce the interlayer slip in $PdSe_2$. Both uniaxial strain and shear strain can induce ferroelastic switching with phase transition, but one only needs to apply 8% shear strain to trigger ferroelastic switching, much smaller than that for uniaxial strain case which needs about 25% strain. Shear strain can also induce so-called three-states ferroelastic switching [34] while uniaxial strain cannot. This will be discussed in details later.

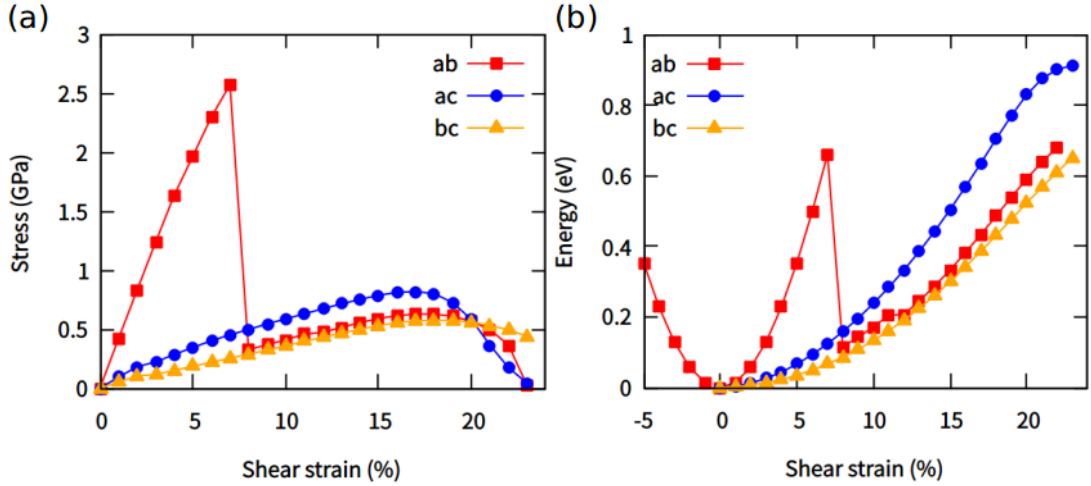

**Figure 2**. (a) The shear stress-strain and (b) corresponding energy-strain relationship curves for bulk PdSe$_2$ with shear strain in *ab* plane and *ac*, *bc* plane, respectively.

We then examined the structural change for bulk PdSe$_2$ under shear strain in *ab* plane. The corresponding lattice constants *b* and *c* also have a sharp change with the increase of lattice constant *b* and the decrease of *c* (**Figure 3b**) when shear strain is at 8%, indicating that the PdSe$_2$ transforms into a new structure. In order to better understand the structural rearrangement, we summarized the bond lengths of PdSe$_2$ under different shear strains in *ab* plane, as shown in **Figure 3c**. With shear strain increases from 0% to 7%, the atoms of Se$_4$ and Pd ($d_4$) in two different layers move together while the distances of Se$_1$-Pd ($d_1$), Se$_2$-Pd ($d_2$), and Se$_1$-Se$_3$ ($d_3$) atoms change slightly. When shear strain increases to 8%, the intralayer bond of Se$_2$-Pd suddenly breaks while a new bond ($d_4$) between Se$_4$ and Pd forms, corresponding to the structural phase transition in **Figure 2**. These results also suggest the vdW layered direction changes from *c* to *b* direction.

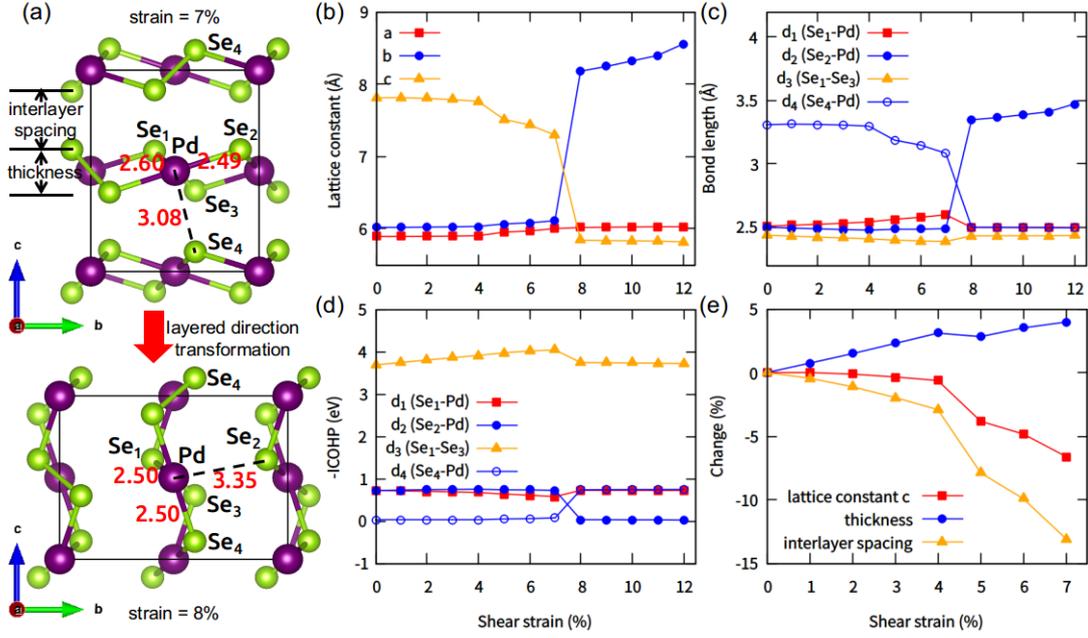

**Figure 3**. (a) The structural rearrangement when applying the shear strain in *ab* plane. With shear strain changing from 7% to 8%: $Se_2$-Pd bond breakage (covalent bond to vdW interaction) while $Se_4$-Pd bond formation (vdW interaction to covalent bond). The red texts are the bond length in Å. The corresponding (b) lattice constant, (c) bond lengths, (d) -ICOHP (eV) evolution. (e) The change (%) of lattice constant *c*, layer thickness and interlayer spacing for bulk $PdSe_2$ under shear strain in *ab* plane. The layer thickness is defined as the difference of the top and bottom atoms in the same layer and interlayer spacing represents the distance of two nearest layers in bulk $PdSe_2$, which are marked in **Figure 3a**.

To confirm the vdW layered direction transformation with bond breakage and formation, we also calculated the integrated crystal orbital Hamilton population (-ICOHP), which indirectly reflects the bond strength in compounds by counting the energy-weighted population of wave functions between two atomic orbitals, as well as the interatomic interactions from electronic perspective [47, 48], as shown in **Figure 3d**. In the strain-free $PdSe_2$ structure, the -ICOHP value between $Se_4$ and Pd in different layers is only 0.03 eV (typical vdW interaction) while other bonds have relatively large value (chemical bonding), in which the -ICOHP value of Se-dimer bond is abnormally large with 3.69 eV (strong covalent interaction). From shear strain of 7% to 8%, the -ICOHP of $Se_4$ and Pd ($d_2$) suddenly increases to 0.74 eV while the -ICOHP of $Se_2$ and Pd decreases to 0.03 eV, suggesting that the $Se_4$ and Pd become chemical bonding from original vdW interaction (out-of-plane to in-plane) while $Se_2$ and Pd bonding breaks. These intralayer bond breakage and interlayer bond formation confirm the vdW layered direction transformation from *c* to *b* direction.

The ultra-strong interaction of $Se_1$ and $Se_3$ makes Se-dimer nearly unchanged and Se-dimers are mainly responsible for the structural stability through the shear strain induced structural rearrangement. By normalizing second order force constant on the unit vector along each bonding direction, we obtained the bond-projected force constant [49, 50] (**Table 1**) to compare the interatomic strength and bonding stiffness of PdSe$_2$. We found that the bond-projected force constant value of Se$_1$-Se$_3$ (7.07 eV/Å$^2$) in Se-dimer is greatly larger than those of the intralayer Se-Pd (4.18 and 4.15 eV/Å$^2$) and interlayer Se$_4$-Pd (0.14 eV/Å$^2$) interactions. Under the shear strain field, Se-dimer still remains ultra-strong strength to avoid structure failure, which is response for the structure stability of PdSe$_2$ during structural rearrangement.

**Table 1**. The bond-projected force constant $\Phi$ (eV/Å$^2$) for four atomic interactions of bulk PdSe$_2$ under different shear strains in *ab* plane.

|    | Se$_1$-Pd | Se$_2$-Pd | Se$_1$-Se$_3$ | Se$_4$-Pd |
|----|-----------|-----------|---------------|-----------|
| 0% | 4.18      | 4.15      | 7.07          | 0.14      |
| 7% | 2.35      | 4.32      | 8.10          | 0.17      |
| 8% | 4.25      | 0.17      | 7.44          | 4.51      |

To understand the layered direction transformation, we defined two quantities: layer thickness and interlayer spacing in the structure of bulk PdSe$_2$, as shown in **Figure 3a**. The values of the change (%) of lattice constant *c* under shear strain in *ab* plane were calculated and shown in **Figure 3e**. It can be seen that the layer thickness obviously increases and the interlayer spacing decreases with shear strain from 0 to 4%, resulting in nearly unchanged lattice constant *c*. As shear strain increasing from 5% to 7%, it shows the more obvious tendency of single-layer expansion and interlayer spacing reduction. With the strain increases, the interlayer Se$_4$ and Pd moves together and forms the bond along *c* direction, which becomes the non-vdW direction in the new phase after ferroelastic transition. While Se$_2$-Pd bond along b direction increases and breaks with shear strain, and finally becomes vdW direction in the new phase.

Usually, the strain driven structural rearrangement is accompanied by the electronic structure change. We calculated the band gap with a function of shear strain and found that shear strain in *ab* plane can firstly reduce the band gap to zero at 4% strain, and then reopen the band gap at 8% and then it continuously increases with shear strain (**Figure 4a**), suggesting strong coupling with interlayer spacing of PdSe$_2$ due to strong interlayer interaction [29, 51]. This semiconductor-metal-semiconductor transition is selective and only occurs in *ab* shearing plane rather than *ac* or *bc* shearing plane, because in later case, shear strain only induces the interlayer slip without significantly electric band structure change. The three detailed band structures for PdSe$_2$ under shear strain in *ab* plane are shown in **Figures 4b-d**. The band structure for strain-free PdSe$_2$ is shown in **Figure 4a** with the gap of 0.06 eV, which well matches the experimental result from dI/dV spectrum [29]. After the semiconductor-metal transition at shear strain of 4%, the PdSe$_2$ continues to possess the 0 eV band gap arising from the decreased interlayer spacing, which appears at that of one band dropping down along S-Y direction and another band rising up to conduction band area near Gamma point at shear strain of 7%. As shear strain increases to 8%, the conduction band minimum (CBM) returns to the original position and valence band maximum (VBM) moves downward, resulting in the reopening of band gap to 0.47 eV. This metal-semiconductor transition responds for the structural rearrangement with interlayer spacing variation and finally new vdW direction formation by the shear strain at shear strain of 8% in *ab* plane.

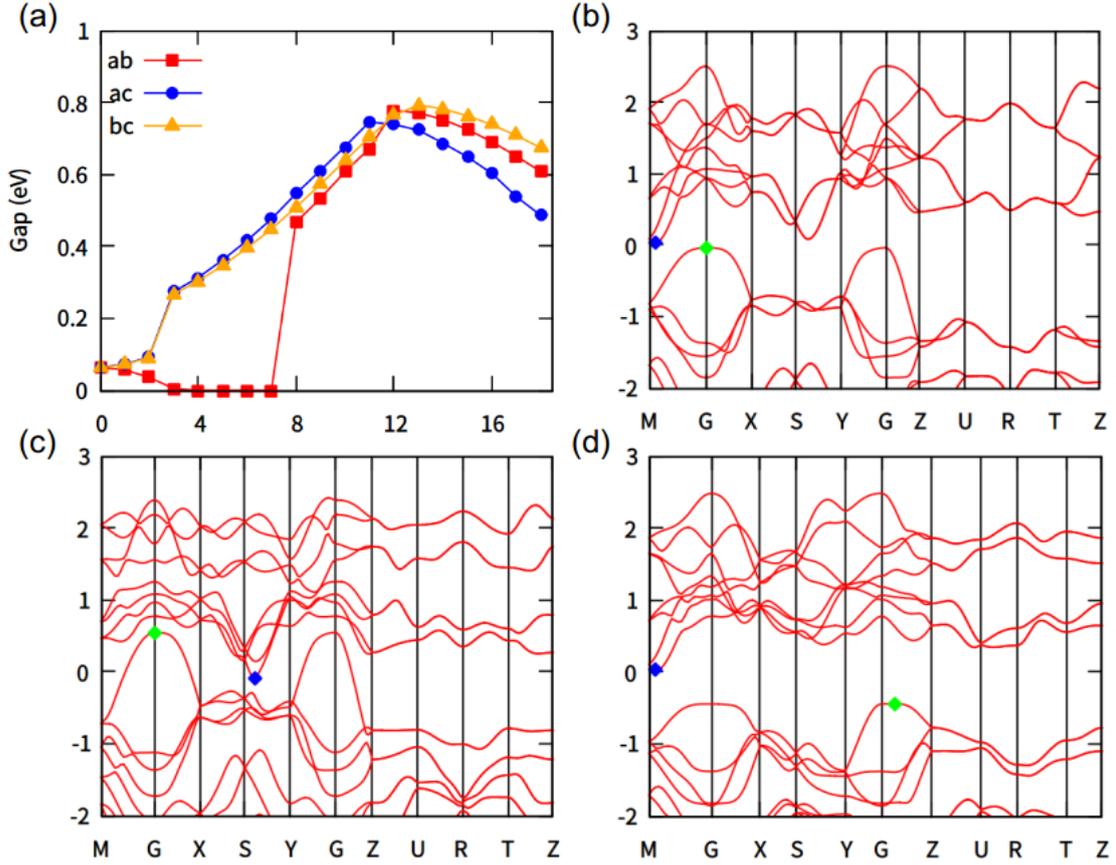

**Figure 4**. (a) The band gap (eV) evolution under shear strain in three planes. The band structures of bulk PdSe$_2$ under shear strain (b) 0%, (c) 7%, and (d) 8% in *ab* plane. The energy scales of band are aligned with respect to the Se-5*s* states. The CBM and VBM are marked as blue and green points, respectively.

Finally, we explore the three-states ferroelastic switching [34] under shear strain in PdSe$_2$. The three-states ferroelastic switching [34] is driven from different ferroelastic variants and has been realized by using band-excitation piezoresponse force microscopy [34]. Interestingly, shear strain can also induce three-states ferroelastic swiching in PdSe$_2$. After the first lattice rotation with vdW direction transform to *b* from *c* direction, if applying shear strain in two steps flowing by *ac* plane and then *bc* plane, one can make PdSe$_2$ return to its initial ferroelastic state, as shown from **Figure 5a-5c**. The corresponding relative energy evolution under shear strain (in-plane) was calculated and shown in **Figure 5d**. It can be seen that the energy needed to overcome for this three-states switching is relatively small, indicating the possibility for its application. Meanwhile, combing the semiconductor-metal-semiconductor transition during this three-states switching, it is expected that PdSe$_2$ may show more interesting functionalities under shear strain. The development of multiple-states ferroelastic switching will inspire people to explore the ferroelastic domain architectures capable of exhibiting new functionalities and improved performance.

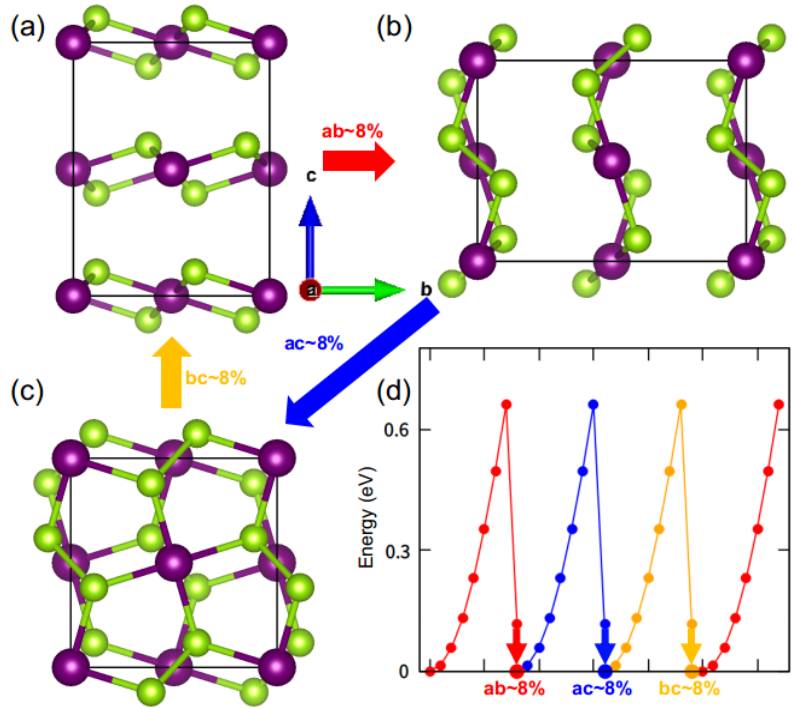

**Figure 5**. (a-c) The three-states ferroelastic switching with different shear strain and (d) corresponding relative energy evolution with in-plane shear strain for bulk $PdSe_2$.

## 4. Conclusions

In summary, we systemically investigated the structure and electronic properties evolution of two-dimensional $PdSe_2$ under uniaxial and shear strains from first-principles method. Our results show that, uniaxial strain along crystal *a*, *b* or *c* directions and in-plane shear strain can induce the ferroelastic switching in $PdSe_2$. The strain induced ferroelastic switching in $PdSe_2$ is accompanied by a lattice rotation with the transformation from non-vdW direction to vdW direction, as well as the semiconductor-metal-semiconductor transition. In addition, the shear strain shows more efficient to induce ferroelastic switching with two times amplitude smaller than the uniaxial strain. Interestingly, the novel three-states ferroelastic switching in two dimensional materials $PdSe_2$ also occurs under shear strain but not through uniaxial strain. Our result shows that the shear strain could be used as an effective approach for manipulating the functionalities of layered materials in potential device applications.

## Acknowledgements

This work is supported by the National Science Foundation of China (Grant No. 11572040) and the Thousand Young Talents Program of China. X.W. acknowledges National Key Research and Development Program of China (2019YFA0307900), Beijing Natural Science Foundation (Grant No. Z190011) and Beijing Institute of Technology Research Fund Program for Young Scholars. Y.L. acknowledges National Natural Science Foundation of China with Grant Nos. 11572040, 11804023 and the China Postdoctoral Science Foundation with Grant No. 2018M641205. Theoretical calculations were performed using resources of the National Supercomputer Centre in Guangzhou, which is supported by Special Program for Applied Research on Super Computation of the NSFC-Guangdong Joint Fund (the second phase) under Grant No. U1501501.

**Supporting information** The Supporting information available online provides additional tables and figures.

# Supporting information for "Ferroelastic switching with van der Waals direction transformation in layered PdSe$_2$ driven by uniaxial and shear strain"

Peng Lv[1], Gang Tang[1], Yanyu Liu[1], Yingzhuo Lun[1], Xueyun Wang[1] and Jiawang Hong[1]*

[1]*School of Aerospace Engineering, Beijing Institute of Technology, Beijing 100081, China*

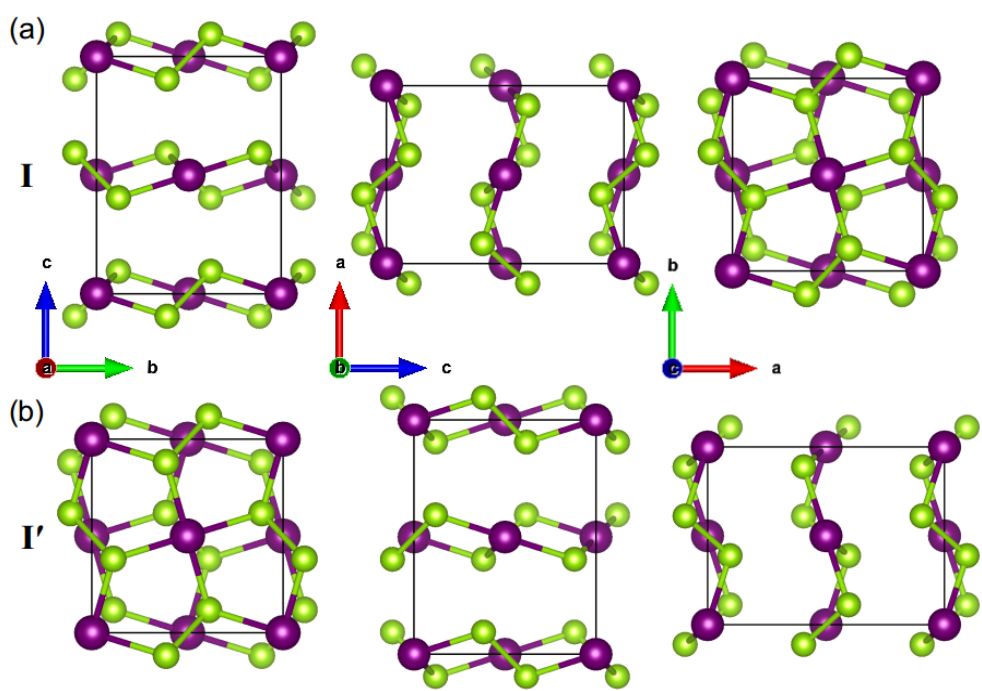

**Figure S1**. The atomic structures of bulk PdSe$_2$ (lattice constants: $a=b'$, $b=c'$, $c=a'$) correspond to the state of (a) I and (b) I′ for uniaxial strain along *a* direction in **Figure 1**. Hereafter, the purple and green spheres denote the Pd and Se atoms, respectively.

* Corresponding author. E-mail: hongjw@bit.edu.cn (J. Hong);

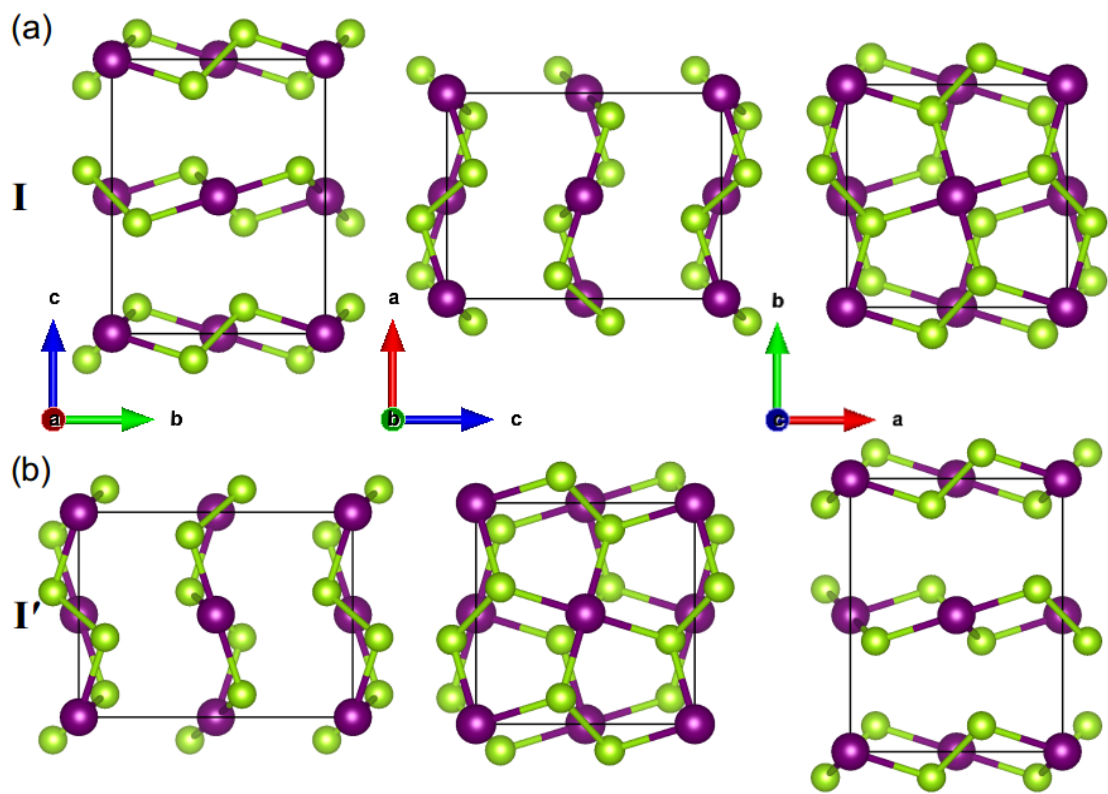

**Figure S2**. The atomic structures of bulk PdSe$_2$ (lattice constants: $a=c'$, $b=a'$, $c=b'$) correspond to the state of (a) I and (b) I′ for uniaxial strain along $b$ direction in **Figure 1**.

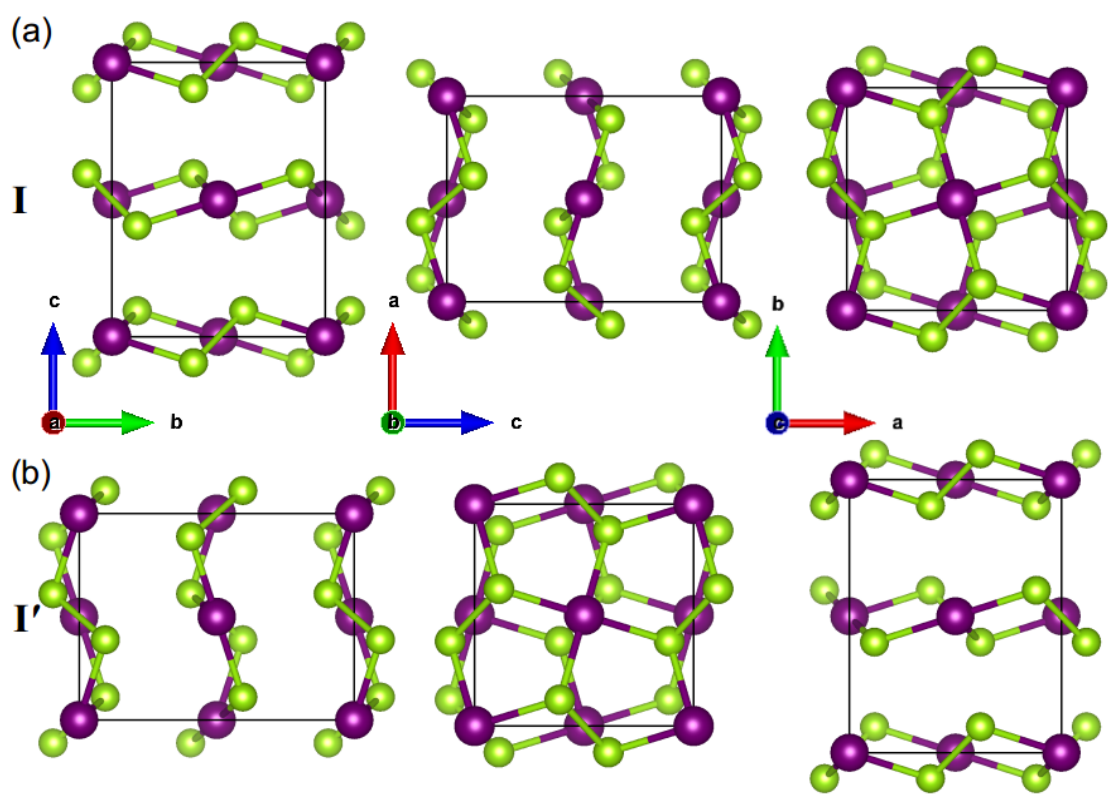

**Figure S3**. The atomic structures of bulk PdSe$_2$ (lattice constants: $a=c'$, $b=a'$, $c=b'$) correspond

to the state of (a) I and (b) I' for uniaxial strain along *c* direction in **Figure 1**.

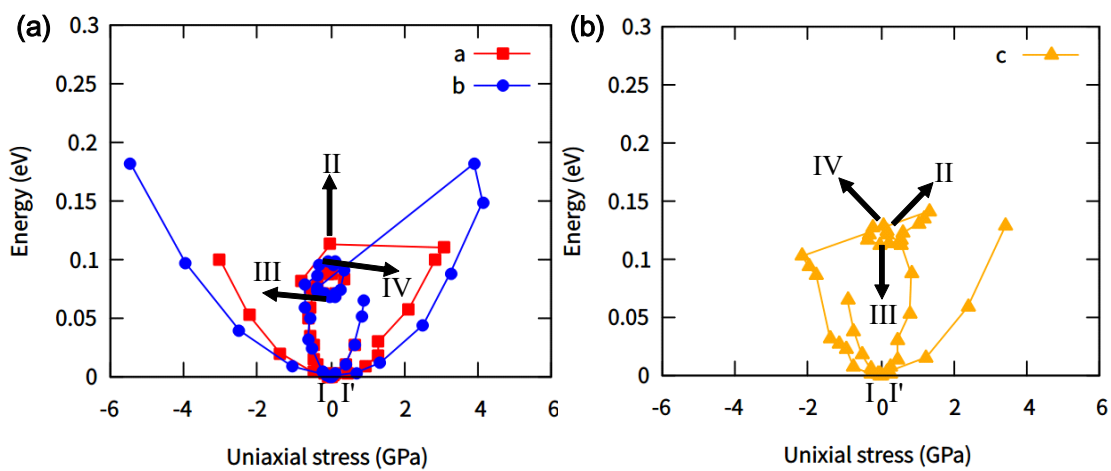

**Figure S4**. The relative energy (eV) with a function of uniaxial stress for bulk PdSe$_2$ along (a) *a*, *b*, (tensile) and (b) *c* (compressive) directions.

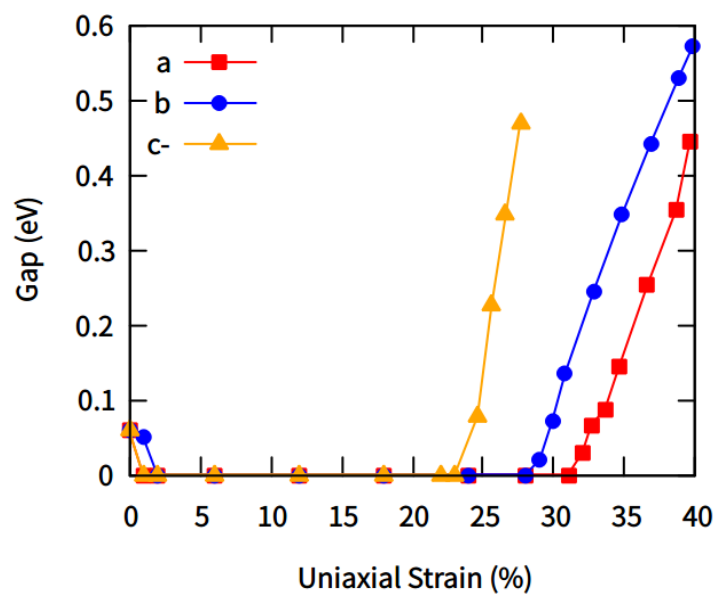

**Figure S5**. The band gap (eV) evolution under uniaxial strain for bulk PdSe$_2$ along *a*, *b*, (tensile) and *c* (compressive) directions.

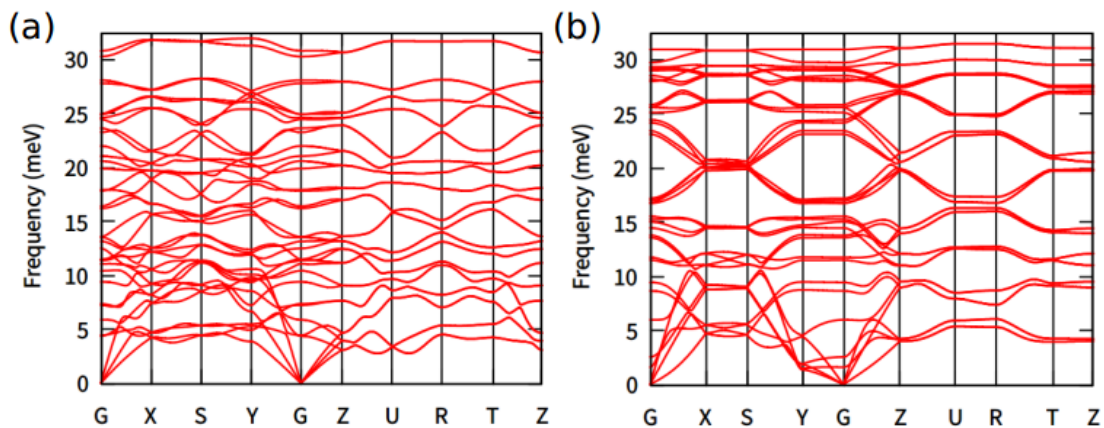

**Figure S6**. The phonon spectrum of bulk PdSe$_2$ under shear strain of (a) 7% and (b) 8% in *ab* plane, respectively.

**Table S1**. The lattice constants (Å) of bulk PdSe$_2$ by using different vdW correction method. For comparison, the values from experiment and PBE are also shown.

| Lattice(Å) | Exp. value[1] | **optPBE** | optB88 | vdW-DF2 | optB86b | PBE |
|---|---|---|---|---|---|---|
| *a* | 5.741 | **5.89** | 5.91 | 6.03 | 6.23 | 5.80 |
| *b* | 5.866 | **6.01** | 6.01 | 6.16 | 6.24 | 5.95 |
| *c* | 7.691 | **7.81** | 7.391 | 8.27 | 6.23 | 8.45 |
| Maximum error | ~ | **2.6%** | 3.9% | 7.5% | 19.0% | 10.3% |

[1] *Journal of the American Chemical Society* **2017**, 139, 14090-14097

**Table S2**. The lattice constants (Å) of bulk PdSe$_2$ correspond to the stable and meta-stable states under uniaxial strain in **Figure 1**:

| Uniaxial strain direction | Lattice | I | III | IV | I' |
|---|---|---|---|---|---|
| Along *a* direction | *a* | 5.89 | 6.36 | 6.83 | 7.81 |
|  | *b* | 6.01 | 6.36 | 6.12 | 5.89 |
|  | *c* | 7.81 | 6.25 | 6.14 | 6.01 |
| Along *b* direction | *a* | 5.89 | 6.35 | 6.18 | 6.01 |
|  | *b* | 6.01 | 6.35 | 6.80 | 7.81 |
|  | *c* | 7.81 | 6.26 | 6.11 | 5.89 |
| Along *c* direction | *a* | 5.89 | 6.32 | 6.17 | 6.01 |
|  | *b* | 6.01 | 6.32 | 6.21 | 7.81 |
|  | *c* | 7.81 | 6.32 | 6.64 | 5.89 |

**Table S3.** Mechanical properties of PdSe$_2$: the elastic constant $C_{ij}$ (GPa), bulk modulus $B$ (GPa), shear modulus $G$ (GPa), Young's modulus $E$ (GPa), Poisson's ratio $v$, and ductility index $B/G$.

| $C_{11}$ | $C_{22}$ | $C_{33}$ | $C_{12}$ | $C_{13}$ | $C_{23}$ | $C_{44}$ | $C_{55}$ | $C_{66}$ |
|---|---|---|---|---|---|---|---|---|
| 126.3 | 157.0 | 19.7 | 29.1 | 29.6 | 23.9 | 5.1 | 9.4 | 41.9 |
| $B$ | $G$ | $E$ | $v$ | $B/G$ | $G_{44}$ | $G_{55}$ | $G_{66}$ | |
| 52.0 | 26.0 | 45.8 | 0.3 | 2.0 | 5.2 | 9.1 | 41.7 | |